# Room-Temperature Noncolinear Ferroelectricity in van der Waals $WO_2Cl_2$ with a Wide Bandgap

Yu Xing[1,#], Ning Ding[1,#], Zhipeng Wang[2,#], Zhiwen Pan[1,#], Lei Guo[1], Guowei Du[3], Yangrui Liu[4], Xiaoxing Cao[5], Ran Su[6], Mengting Jiang[7], Xuezhi Ma[7], Xiyu Chen[1], Junchao Zhang[1], Xinyu Yang[1], Haoran Ye[1], Honghong Yao[1], Rui Feng[1], Dexiang Chen[1], Le-Ping Miao[8], Yumeng You[5], Zejun Li[1], Dongsheng Song[4,*], Linglong Li[1,*] and Shuai Dong[1,*]

[1]Key Laboratory of Quantum Materials and Devices of Ministry of Education, School of Physics, Southeast University, Nanjing 211189, China

[2] School of Advanced Materials and Nanotechnology, Xidian University, Xi'an 710126, China

[3]Jiangsu Key Laboratory of Advanced Laser Materials and Devices, School of Physics and Electronic Engineering, Jiangsu Normal University, Xuzhou 221116, China

[4]Information Materials and Intelligent Sensing Laboratory of Anhui Province, Institutes of Physical Science and Information Technology, Anhui University, Hefei 230601, China

[5]Jiangsu Key Laboratory for Science and Applications of Molecular Ferroelectrics, Southeast University, Nanjing 211189, China

[6]College of Science, Hebei University of Science and Technology, Shijiazhuang 050018, China

[7]Quantum Innovation Centre (Q. InC), Agency for Science, Technology and Research (A*STAR), 4 Fusionopolis Way, Kinesis, #05, Singapore, 138635, Republic of Singapore

[8]Chaotic Matter Science Research Center, Department of Materials, Metallurgy and Chemistry & Jiangxi Provincial Key Laboratory of Functional Molecular Materials Chemistry, Jiangxi University of Science and Technology, Ganzhou 341099, China

# These authors contributed equally to this work.

* Corresponding authors: dsong@ahu.edu.cn; linglongli@seu.edu.cn; sdong@seu.edu.cn

**Abstract:** Low-dimensional ferroelectrics are attractive for their promising prospects in nanoelectronics. Compared with widely-used ferroelectric perovskites, most low-dimensional ferroelectrics exhibit several inborn weaknesses such as small bandgaps (mostly <2 eV, i.e. semiconductors-like) or faint polarizations (e.g. <1 μC/cm$^2$ for sliding ferroelectrics even if their bandgaps can be large). Here we experimentally demonstrate the room-temperature ferroelectricity of van der Waals $WO_2Cl_2$. The well-tested $d^0$ rule inherited from ferroelectric perovskites leads to a large dipole (~3 eÅ) from the off-center displacement of $W^{6+}$ ion and a wide bandgap of 2.80 eV. Its ferroelectricity is proved by multiple characterizations including second harmonic generation, piezoresponse force microscopy, and ferroelectric hysteresis loops. More interestingly, the exotic noncollinear dipole order is directly observed at the atomic level by integrated differential phase contrast scanning transmission electron microscopy. Our work paves an alternative route for low-dimensional ferroelectrics to pursue excellent ferroelectric performance and distinct physics of polarity.

Two-dimensional (2D) ferroelectric materials exhibit spontaneous polarizations in a single or few layers, making them promising candidates for ultra-thin memory and nano-electronic devices. In contrast to conventional ferroelectric perovskites with three-dimensional (3D) frameworks, many 2D ferroelectrics are free from dangling bonds at surface/interface, thus they can be largely immune to depolarization fields and surface/interfacial effects, allowing robust polarizations to persist down to the ultra-thin limit. The experimentally confirmed 2D ferroelectrics include $CuInP_2S_6$ [1, 2], $In_2Se_3$ [3, 4], $WTe_2$ [5, 6], *MX* (*M*=Sn, Ge; *X*=S, Se, Te) [7-11], $NbOX_2$ (*X*=I, Cl) [12-15], as well as many sliding ferroelectrics like *h*-BN bilayer and others [16-19].

For most of them, the origins of their ferroelectric polarizations are improper, with novel mechanisms beyond the common rules for 3D perovskites. However, a concomitant weakness is their narrow bandgaps (mostly <2 eV), or in other words many of them are more like semiconductors instead of dielectrics. Consequently, under applied electrical fields, large leakage currents are unavoidable, making the ferroelectric switching challenging and unwished Joule heat. Another weakness of these 2D improper ferroelectrics is that their polarizations are typically small (mostly $<10\ \mu C/cm^2$). Even if the bandgap issue can be solved in some sliding ferroelectrics based on highly insulating layers like BN and hybrid molecular [16, 20], their polarizations are too faint, e.g. $<1\ \mu C/cm^2$, which are incompetent for broad applications. The only exceptional member is $CuInP_2S_6$, which owns a moderate polarization of 2.55 $\mu C/cm^2$ and a large bandgap of 2.7 eV [21]. However, its ferroelectric Curie temperature ($T_C$) is only 315 K and its ionic conductance is prominent [22]. Thus, it is urgent to pursue more room-temperature 2D ferroelectrics with wider bandgaps and larger polarizations, which are vital for advancing next-generation devices based on 2D ferroelectrics.

Different from these improper ferroelectrics, the transition metal dioxydihalides $MO_2X_2$ (*M*=W, Mo; *X*=Cl, Br), a class of van der Waals (vdW) material, was predicted to own proper ferroelectricity [23]. With a perovskite-like structure, their ferroelectricity originated from the well-tested $d^0$ rule as in the well-established Ti-based ferroelectric perovskites. Taking $WO_2Cl_2$ for example, the $W^{6+}$ ion with empty 5*d* orbitals tends to form coordination bonds with neighboring O's 2*p* orbitals, leading to the off-center displacement in the $O_4Cl_2$ octahedron. A relatively large bandgap of ~2-3 eV is expected according to *ab initio* calculations [24-26]. More interestingly, the competition between a ferroelectric phonon mode and an antiferroelectric phonon mode in $WO_2Cl_2$ led to a scenario similar to magnetic frustration. As a result, a noncollinear order of dipoles was predicted (coined as noncollinear ferrielectricity in Ref. [23]), with compensated dipoles along one W-O axis and uncompensated dipoles along another W-O axis. Even though, the strong coordination-bonding interaction and high valence of $W^{6+}$ lead to a large residual spontaneous polarization (~26 μC/cm$^2$) [23], comparable to that of $BaTiO_3$ (20~25 μC/cm$^2$). Despite these pioneer theoretical predictions, direct experimental evidence of ferroelectricity in $WO_2Cl_2$ remains elusive, mainly due to its ambient instability.

Furthermore, the noncollinear electric dipole textures from phonon mode frustration are also crucial but extremely challenging for experimental observation since it is on the atomic scale. In fact, the noncollinear magnetic orders, such as cycloids [27], helices [28, 29], and skyrmions [30], have become one core spot in condensed matter. However, the ferroelectric counterparts, i.e. noncollinear dipole textures, remain exceptionally rare and poorly understood. Currently, the experimental observation of dipole vortices and skyrmions was limited in $PbTiO_3/SrTiO_3$ superlattices, generated via classical electrostatic boundary conditions and usually in the mesoscale [31]. Although some molecular ferroelectrics can naturally exhibit noncollinear dipoles due to the different orientations of polar groups [32, 33], their noncollinearity is somehow trivial due to the geometric locking, i.e. unadjustability. There is only one experimental pioneer work reporting the intrinsic noncollinear dipole order in $BiCu_{0.1}Mn_{6.9}O_{12}$ via neutron diffraction analysis [34], which is also driven by phonon mode frustration.

Here, we report the experimental demonstration of room-temperature noncolinear ferroelectricity in $WO_2Cl_2$. Polarization-resolved Raman and second-harmonic generation (SHG) spectroscopy further indicates pronounced in-plane anisotropy, supporting the presence of directional dipolar ordering. Crucially, piezoresponse force microscopy (PFM) reveals alternating striped ferroelectric domains and writable domains. Most importantly, the standard ferroelectric loops directly confirm the ferroelectricity of $WO_2Cl_2$. Furthermore, using the integrated differential phase contrast (iDPC) scanning transmission electron microscopy (STEM), the noncollinear dipole order in the atomic scale is directly visualized.

$WO_2Cl_2$ single crystals are synthesized using the chemical vapor transport (CVT) method. The crystals are transparent and nearly colorless [inset of Fig. 1(a)], which can be exfoliated into few-layer flakes. As shown in Fig. 1(a), the X-ray diffraction (XRD) pattern displays sharp, well-defined peaks characteristic of single-crystal, indicating high crystallinity. Complementary powder XRD measurements further confirm that the material crystallizes in the non-centrosymmetric *Cc* space group (No. 9), with structural parameters refined via Rietveld analysis, as presented in Fig. S1 and Table S1 in Supplemental Material (SM) [35]. Moreover, electron energy dispersive spectroscopy (EDS) mapping images reveal a uniform distribution of W, O, and Cl elements throughout the sample [Fig. S2 in SM [35]], verifying the compositional homogeneity.

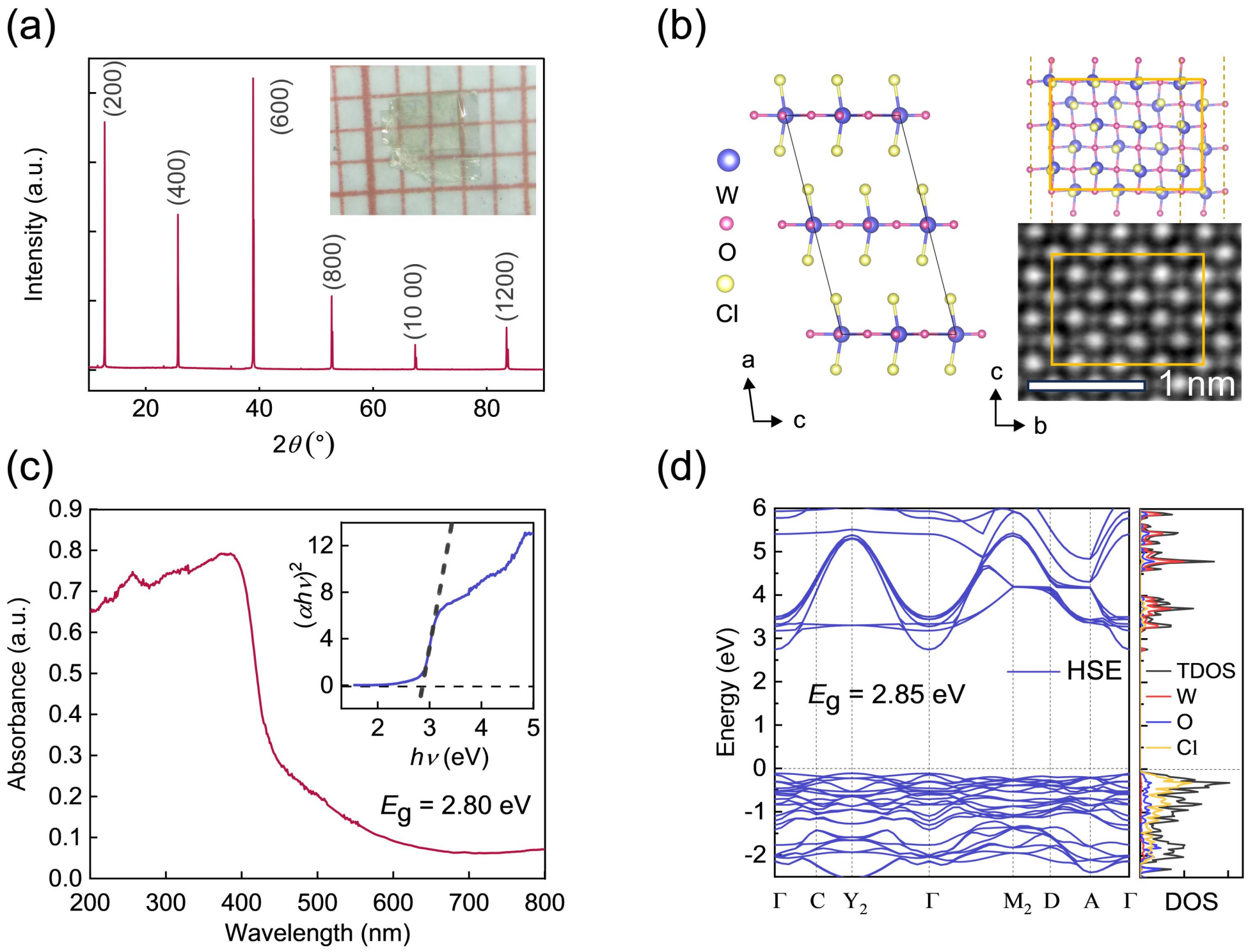

**FIG. 1.** Crystalline and electronic structures of $WO_2Cl_2$. (a) Single-crystal XRD pattern. Inset: the as-grown optically transparent crystal. (b) The A-B stacking vdW layers. Left: side view. Right: top view, in comparison with iDPC-STEM image. (c) Ultraviolet-visible (UV-Vis) absorption spectrum. Inset: the Tauc-plot with fitting, indicating a bandgap of 2.80 eV. (d) Band structure and density of states calculated by the Heyd-Scuseria-Ernzerhof (HSE06) hybrid functional method.

As a vdW dioxydihalide, its layered structure is illustrated in Fig. 1(b). These octahedra connect in the corner-sharing mode, forming a quasi-square lattice in the *bc* plane. The stacking mode is the A-B type, instead of the A-A one (space group $P2_1am$, No. 26) as mentioned in some previous theoretical studies [36, 37]. The electron diffraction pattern of the $WO_2Cl_2$ flake, as shown in Fig. S3 in SM [35], confirms its high crystalline quality. The simulated diffraction pattern based on the atomic model in Fig. 1(b) matches well with the experimental results, verifying that the normal vector of the $WO_2Cl_2$ flake is aligned with the normal of *bc* plane. When viewed along this direction, the structure exhibits an apparent overlap of Cl and W ions, while the O ion is located between two neighboring $WCl_2$ columns along both *b* and *c* axes [38-40]. Such A-B stacking configuration of $WO_2Cl_2$ is further confirmed by STEM. High-resolution iDPC-STEM imaging reveals that the bottom layer is laterally shifted along the [011] direction in the *bc* plane, thus the W ions in the lower layer occupy the central vacancies defined by four W ions in the upper layer.

The bandgap of $WO_2Cl_2$ is experimentally determined to be 2.80 eV from the ultraviolet-visible absorption spectrum [Fig. 1(c)], according to the Tauc equation $(\alpha h\nu)^n = A(h\nu - E_g)$ [41]. Density functional theory (DFT) calculations are performed using the hybrid functional [25]. The theoretical bandgap is 2.85 eV [Fig. 1(d)], remarkably consistent with the experimental one. Projected density of states (PDOS) analysis indicates that the valence band maximum is primarily derived from Cl's 3*p* orbitals, while the conduction band minimum originates mainly from W's 5*d* orbitals. The PDOS supports the $d^0$ scenario for $W^{6+}$, which makes $WO_2Cl_2$ a good band insulator. The large bandgap endows $WO_2Cl_2$ with excellent insulating properties, high thermal stability, and potential for applications in ultraviolet optoelectronic and high-power electronic devices [42].

As predicted in Ref. [23], the emerging dipoles from $d^0$ $W^{6+}$ ions will break the $C_4$ rotation symmetry of the in-plane square lattice. To investigate the in-plane anisotropy of $WO_2Cl_2$, angle-resolved Raman and SHG spectroscopy measurements are performed at room temperature under vacuum conditions. As shown in Fig. 2(a), the Raman spectrum reveals seven distinct vibrational modes for $WO_2Cl_2$. Polar plots of the Raman intensity at 155.4 $cm^{-1}$ under parallel and perpendicular polarizations are presented in Figs. 2(b-c), respectively. Here, the parallel and perpendicular configurations refer to whether the polarizer and analyzer are aligned or orthogonal to each other. Theoretical fittings derived from Raman intensity equations show excellent agreement with the experimental data (see Fig. S4 in SM [35]). Additional angle-resolved Raman measurements are presented in Fig. S5 in SM [35], further supporting the in-plane vibrational anisotropy of the $WO_2Cl_2$ crystal.

SHG serves as a powerful, non-destructive probe for detecting non-centrosymmetric crystal structures. When a sample is illuminated with light of fundamental frequency $\omega$, emission light at the second harmonic frequency $2\omega$ can be detected [43]. The power-dependent SHG intensity of $WO_2Cl_2$ is presented in Fig. 2(d), demonstrating a strong SHG response at room temperature, with visible green emission (532 nm) observed under laser excitation (1064 nm). Furthermore, the polar plots in Figs. 2(e-f) show the SHG intensity patterns of $WO_2Cl_2$ under different polarization configurations, where the incident and detected light polarizations are parallel and perpendicular, respectively. A characteristic two-lobed pattern with 180° rotational symmetry is observed in parallel configuration, while a four-lobed pattern is presented in perpendicular configuration. The distinct two-lobed and four-lobed patterns in the two polarization configurations imply a low-symmetry, non-centrosymmetric structure. The observed polarization-dependent anisotropy in both Raman and SHG measurements provides compelling evidence for room-temperature polar anisotropy in $WO_2Cl_2$.

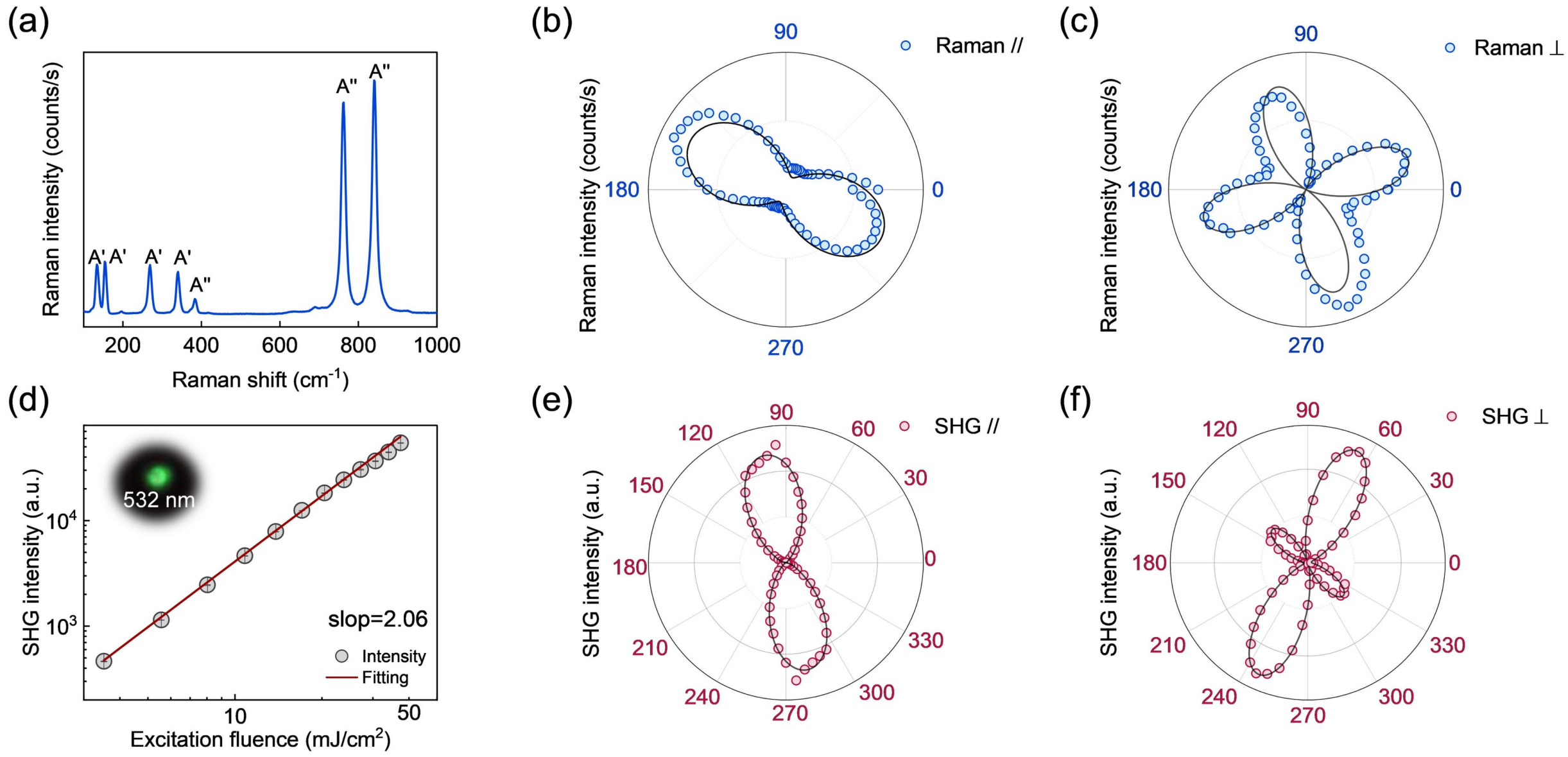

**FIG. 2.** Anisotropic properties of $WO_2Cl_2$ crystals. (a) Raman spectrum of $WO_2Cl_2$ flakes under 532 nm laser excitation. (b-c) Angle-dependent Raman intensity measured under parallel (b) and perpendicular (c) polarization configurations. (d) Log plot of pump fluence dependence of integrated SHG. The slope of the plot is ~2.06, confirming the nature of the SHG emission. Inset: visible green SHG emission (532 nm) generated from the 1064 nm excitation laser. (e-f) Polar plots of SHG intensity under both parallel (e) and perpendicular (f) polarization configurations. Black curves indicate the theoretical fits.

A series of temperature-dependent measurements, including SHG, Raman, XRD, as well as dielectric constant, were conducted to determine the possible ferroelectric transition temperature ($T_C$). As shown in Figs. S6 and S7(a) in SM [35], with increasing temperature, the rapid decrease of SHG signal is evidenced since $T^*$~430 K and a broad peak of dielectric constant appears around $T^*$. However, there is no evidence of structural phase transition across $T^*$ according to the XRD and Raman spectra [Figs. S7(b) and S8 in SM [35]]. Thus, this $T^*$ is not the ferroelectric $T_C$, but the starting point for thermal degradation of sample, as discussed in SM [35]. It is reasonable considering the relative weaker metal-halogen bonds, comparing with the strong metal-oxygen bonds in oxides. Thus, the real ferroelectric $T_C$ of $WO_2Cl_2$ should be higher than 430 K, which is unavailable in our experiments.

To directly visualize the ferroelectric domains, $WO_2Cl_2$ flakes are exfoliated onto $SiO_2$ substrates and characterized using the PFM technique. Intrinsic alternating striped ferroelectric domains are observed in $WO_2Cl_2$ flakes at room temperature. In the PFM phase images, contrast reflects the polarization direction, while the amplitude images correspond to polarization magnitude. To investigate the vector characteristics and anisotropy of the ferroelectric domains, the $WO_2Cl_2$ flakes are systematically rotated by 0°, 45°, and 90° and characterized using both in-plane and out-of-plane PFM modes [Figs. 3(a-f)]. The corresponding surface morphology and amplitude maps are shown in Fig. S9 in SM [35]. The sample orientation is defined such that 0° corresponds to the polarization axis being orthogonal to the long axis of the cantilever. Upon 90° rotation, the in-plane domain contrast markedly diminished, whereas phase contrast from the out-of-plane channel is relatively small, implying a dominant contribution from in-plane polarization. This observation confirms that the spontaneous ferroelectric polarization in $WO_2Cl_2$ predominantly orients within the in-plane direction with 180° ferroelectric domain wall configuration.

Furthermore, local domain writing experiments were performed using a conductive tip. As shown in Fig. 3(g), PFM phase image shows an in-plane box-in-box polarization writing pattern, while the topography and phase images are provided in Fig. S10 in SM [35]. In addition, single-point switching spectroscopic PFM (SSPFM) hysteresis loops display clear 180° phase switching and the characteristic butterfly-shaped amplitude-voltage response, as shown in Fig. 3(h). Both the amplitude and phase loops exhibit a noticeable shift towards positive bias, likely attributable to built-in electric fields present within the thin-film material. Additional PFM images illustrating domain structures at varying flake thicknesses are provided in Fig. S11 in SM [35].

Polarization-electric field (*P*-*E*) hysteresis loops serve as the most credible evidence to verify ferroelectricity. Ideal *P*-*E* loops are generally difficult to acquire for 2D ferroelectrics [12-15], except $CuInP_2S_6$ [1]. The relatively wide band gap of $WO_2Cl_2$ raises the possibility of obtaining standard *P*-*E* hysteresis loops. Nevertheless, these *P*-*E* loops measured at room temperature exhibit distorted shapes due to predominant leakage current. This phenomenon originates from its electrical transport band gap (merely 0.47 eV) [Fig. S12 of SM [35]], which is substantially narrower than the optical band gap of 2.80 eV. After suppressing leakage conduction at low temperatures, well-defined *P*-*E* hysteresis loops can be successfully obtained, as presented in Fig. 3(i) and Fig. S13 in SM [35]. Although the loops are not perfectly square-shaped and the polarization value remains semiquantitative [as discussed in SM [35]], they already rank among the highest-quality results reported for van der Waals oxyhalides to date, to the best of our knowledge.

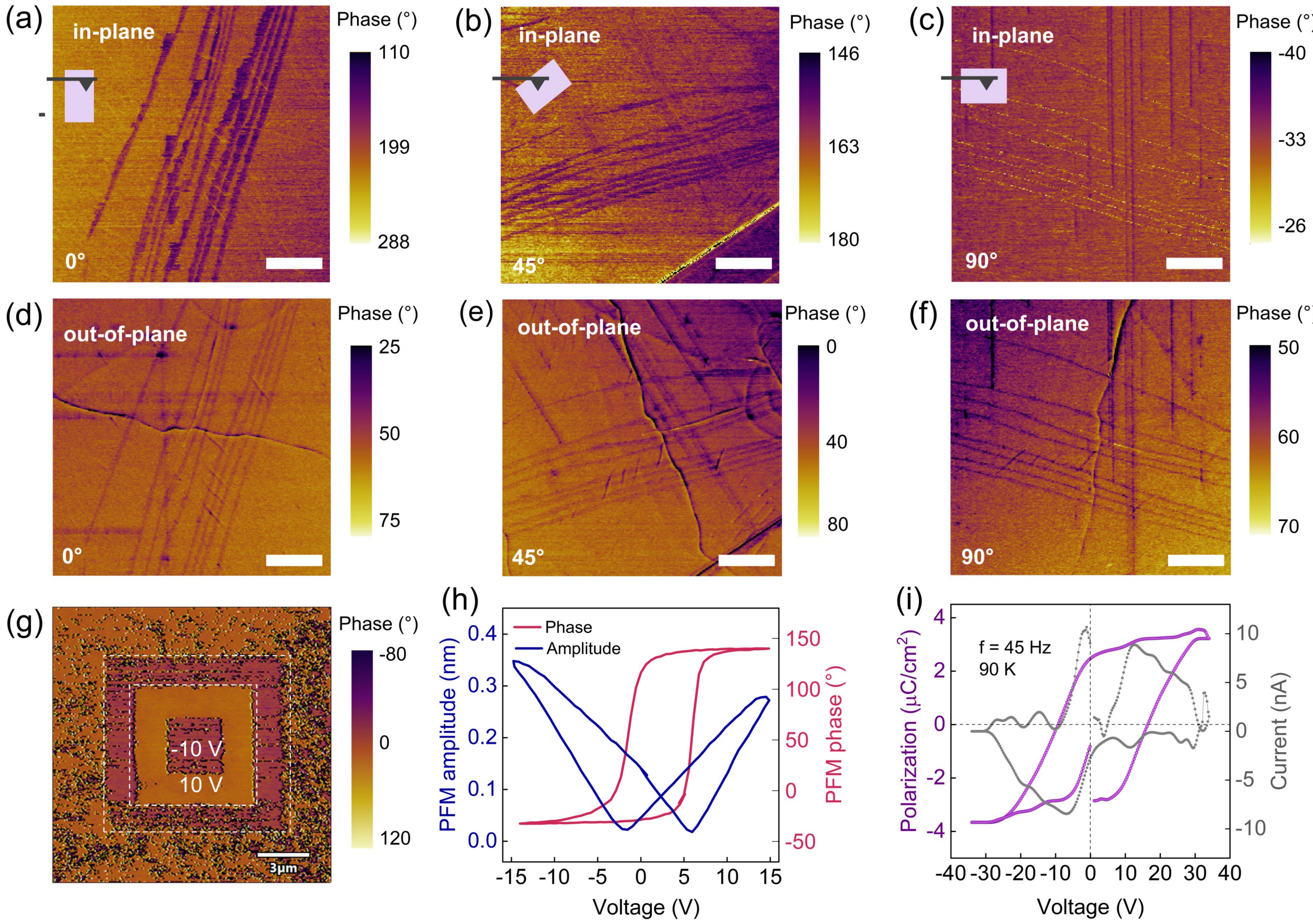

**FIG. 3.** (a-h) PFM characterizations of $WO_2Cl_2$ placed on a $SiO_2$ substrate with a platinum buffer layer. (a-f) In-plane and out-of-plane phase images of a 28 nm nanoflake acquired at rotation angles of 0°, 45°, and 90°, respectively. The 0° orientation is defined such that the polarization axis is orthogonal to the long axis of the PFM cantilever. The scale bar is 2 μm. (g) Box-in-box PFM phase reversal after applying ±10 V square-wave bias. (h) PFM phase and amplitude hysteresis loops measured under 15 V bias. (i) The ferroelectric *P-E* loop measured at 90 K and its corresponding current loop. The measuring frequency is 45 Hz.

The bandgaps, $T_C$ values, and polarizations of experimentally verified 2D ferroelectric materials are summarized in Fig. S14 and Table S2 in SM [35]. For most of them, the bandgaps are not large, corresponding to the absorption of a full-wave band of visible light. Although *h*-BN and (15-crown-5) $Cd_3Cl_6$ own larger band gaps, their polarizations originate from layer sliding, which are rather faint. Notably, the bandgap of $WO_2Cl_2$ is comparable to that of $CuInP_2S_6$ [1, 2, 21], whose absorption edge reaches the purple light. The expected ferroelectric $T_C$ of $WO_2Cl_2$ is higher than that of $CuInP_2S_6$ for more than 100 K, and the polarization is also among the largest ones of 2D ferroelectrics according to the DFT calculation (Fig. S15 and Table S3 in SM [35]).

To further reveal the polarity in the atomic level, we performed the STEM study. Figure 4(a) shows an all-atom-resolved iDPC-STEM image of a $WO_2Cl_2$ crystal sheet. In the enlarged view [Fig. 4(b)], we overlay the projected crystal structure: the larger atomic columns (purple and yellow circles) correspond to overlapping W-Cl atoms, while the smaller columns (pink circles) represent the oxygen sites. Local dipoles in $WO_2Cl_2$ arise from off-center displacements of $W^{6+}$ ions relative to their $O_4Cl_2$ octahedra. We depict the projected W-O-Cl cages as grey rhombuses, with arrows indicating the dipole vectors. These $W^{6+}$ displacements indeed have opposite components along the $b$ direction but identical components along the $c$ direction, supporting intrinsic atomic-level noncollinear ferroelectricity predicted in Ref. [23].

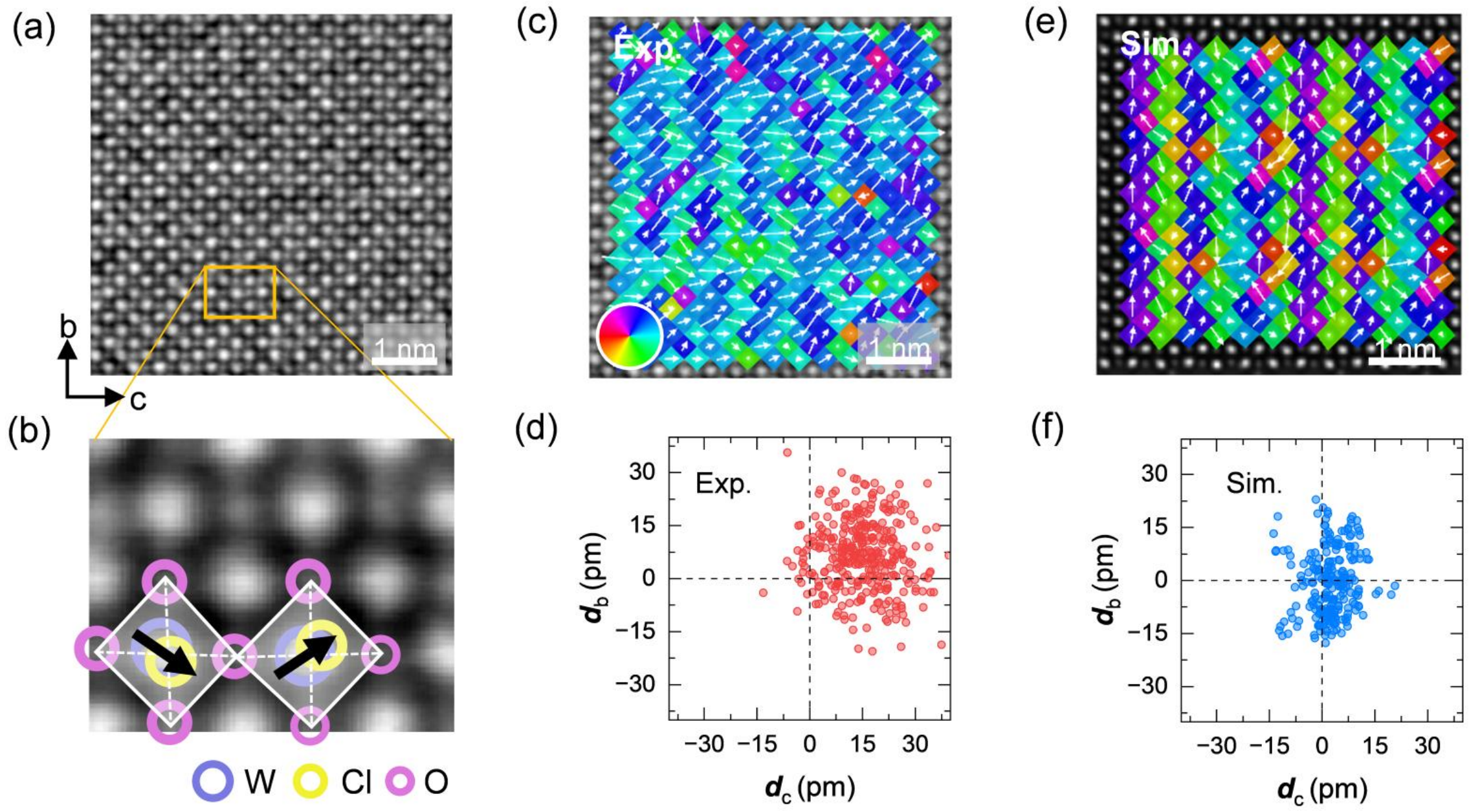

**FIG. 4.** Atomic-scale visualization and analysis of polarization in $WO_2Cl_2$. (a) Atomically resolved iDPC image of $WO_2Cl_2$. The yellow rectangle outlines the region enlarged in (b) showing the A-B stacking. (b) Magnified view overlaid with crystal schematic: gray rhombuses denote projected W-O-Cl octahedra, and black arrows mark the displacement of projected W-Cl columns respected to the centrosymmetric positions defined by the surrounding four oxygen atoms. (c-e) Displacement maps of the projected W-Cl columns derived from (c) experimental image and (e) simulated one at 300 K, with vector directions color-coded at each octahedral site. (d-f) Quantitative projections of the displacement vectors along the $b$ and $c$ crystallographic axes, extracted from (d) experimental and (f) simulated images, respectively.

In Fig. 4(c), we extract local dipole vectors [region #1 in Figs. S16(a-b) in SM [35]] and color-map their orientations. The corresponding statistical projections in Fig. 4(d) confirm a uniform alignment of the $c$-axis components ($d_c$), yielding a mean value of 15.6 pm and a spontaneous polarization along the $c$-axis, while the $b$-axis dipole components ($d_b$) are nearly antiparallel, producing only a small residual moment. The stacking layer-dependent behavior is presented in Fig. S17 in SM [35]. Moreover, the simulated iDPC image from *ab initio* molecular dynamics [Fig. 4(e)] captures the stagger arrangements of dipole strips, and the corresponding statistical analysis in Fig. 4(f) reveals a negligible net dipole along the $b$-axis and a finite value for the $c$-axis component of dipoles.

Further inspection uncovers multiple local dipole configurations within the same crystal. Region #2 shown in Figs. S16(c-d) in SM exhibits alternating strips with a net polarization along the -$c$ axis in both A and B layers [the statistical projections in Figs S18(a-c) in SM [35]], whereas region #3 features oppositely oriented $c$-axis dipoles in the neighboring layers, yielding a cancellation of polarization along the $c$ axis [Fig. S18(d-f) in SM [35]]. These results suggest the weak coupling between neighboring layers at room temperature, which is reasonable for vdW ferroelectrics. Intriguingly, region #4 in Fig. S16 in SM displays a conventional ferroelectric domain with uniformly aligned dipoles across both layers, showing a pronounced net polarization in this material. The corresponding vector ordering is shown in Fig. S18(g-i) in SM [35]. The emerging of a few collinear ferroelectric domains may be driven by thermal excitation, which goes beyond the noncollinear ground state.

In conclusion, we have synthesized and thoroughly characterized $WO_2Cl_2$ as a room-temperature 2D ferroelectric. This material exhibits a wide electronic bandgap, pronounced optical anisotropy, a strong SHG response, and intrinsic striped ferroelectric domains. By employing atomic-scale STEM analysis, we have quantitatively resolved its spontaneous polarization and demonstrated its noncollinear nature at the atomic level. Our findings not only expand the family of known 2D ferroelectrics but also open new avenues for exploration within the dioxydihalides class, particularly in nonlinear optics and ferroelectric field-effect transistor applications.

*Note added.* Recently, we have noticed a relevant work by Fu *et al.* [53], which reported the experimental observation of noncollinear ferrielectricity in an analogous compound $WO_2Br_2$. Their independent findings are consistent with our results.

## Acknowledgments

We thank Profs. Yang Li, Lu You, Chengliang Lu, Meifeng Liu, Xiaoguang Li, Xueyun Wang for the fruitful suggestions with them. We are grateful to Xian-Jiang Song for his assistance with the PFM testing. This work was supported by National Natural Science Foundation of China (Grant Nos. 12325401, 12574091, 12204096, 12504099), the Open Research Funds of Beijing National Laboratory for Condensed Matter Physics (Grant No. 2024BNLCMPKF019), and State Key Laboratory of Advanced Technology for Materials Synthesis and Processing (Wuhan University of Technology). We also acknowledge the computational resources provided by the Big Data Computing Center of Southeast University.

**End Matter**

*Appendix A : Crystal synthesis and structure analysis.*

High-quality $WO_2Cl_2$ single crystals were synthesized using the chemical vapor transport (CVT) method. Stoichiometric amounts of $WCl_6$ and $WO_3$ were mixed in a 1:2 molar ratio inside a quartz tube under an inert atmosphere in a glove box. The mixture was subjected to a thermal gradient of 270 °C (source) to 230 °C (sink) for 7 days, resulting in the crystallization of transparent $WO_2Cl_2$ bulk crystals at the low-temperature end of the tube. XRD measured by X-RAY. EDS was measured using Talos P200X G2. The light absorption characteristics of the samples were obtained using a UV-visible (UV-vis) diffuse reflectance spectrophotometer (Thermo Scientific Evolution 220).

For characterizing anisotropy, Raman spectroscopy and SHG spectra were performed using a WITec alpha 300RA Raman microscope. Raman spectra were acquired with a 532nm laser as the excitation source. SHG measurements were equipped with a 1064 nm pulsed laser with a pulse width of approximately 15 picoseconds and a repetition rate of 80 MHz.

*Appendix B: PFM, P-E hysteresis loop, STM measurements.*

The PFM measurements were carried out using a commercial atomic force microscopy system (Cypher, Asylum Research) housed in a nitrogen-protected glove box. Conductive probes with a nominal spring constant of ~2.8 nN/nm and a resonance frequency of ~75 kHz was used to characterize the domain structure under single-frequency in-plane mode. The amplitude butterfly loop and phase switching loop shown in Fig. 3(h) were obtained in the OFF mode, where the PFM response was recorded after removing the applied bias. And the ferroelectric polarization-electric field (*P*-*E*) hysteresis loops were measured with commercially available ferroelectric systems (Multiferroic Radiant Technologies and Yanhe FETS-2000). A vacuum high-low temperature probe station is used for cooling and sample protection.

For STEM characterization, high-angle annular dark-field STEM (HAADF-STEM) and integrated differential phase contrast STEM (iDPC-STEM) imaging were conducted using a Titan Themis Z microscope operating at an accelerating voltage of 300 kV. DPC-STEM simulations were performed by using a multi-slice frozen phonon model implemented in Dr. Probe [44]. A 300 keV, aberration-free probe with a semi-angle of 25 mrad was assumed. The effective source profile was estimated via Gaussian distribution. The half-width-half-maximum was 0.35 Å. The simulated DPC images were integrated to form iDPC images by custom-written Python code. The atom positions were determined using a two-dimensional Gaussian peak-fitting method implemented in the open-source Python package abTEM [45].

*Appendix C: DFT calculations.*

The first-principles calculations were performed with projector augmented-wave (PAW) potentials as implemented in the Vienna ab initio Simulation Package (VASP) [46]. The Perdew-Burke-Ernzerhof (PBE) parametrization of the generalized gradient approximation (GGA) was used for the exchange-correlation functional [47]. The plane-wave cutoff energy was 500 eV, determined through rigorous convergence testing to ensure computational accuracy. In particular, the W's $5p6s5d$ electrons were treated as valence states. The k-point grid of 3×9×5 was employed for structural relaxation and static computation. To describe the interlayer interaction, two different types of van der Waals corrections of dispersion-corrected density functional theory DFT-D2 and DFT-D3 were used for comparison [48,49]. The convergence criterion of energy was set to $10^{-6}$ eV, and the criterion of Hellman-Feynman forces during the structural relaxation was 0.01 eV/Å. The polarization was calculated by the standard Berry phase method [50,51]. In addition, the Heyd-Scuseria-Ernzerhof (HSE06) functional was employed to address the well-known band-gap underestimation inherent in standard GGA functionals. The ab initio molecular dynamic (AIMD) simulations were performed at 200 K using an NVT ensemble that lasts 5 ps with a time step of 1 fs [52].